\documentclass[linenumbers, manuscript]{aastex631}

\usepackage{amsmath}
\usepackage{soul}

\begin{document}
\begin{nolinenumbers}

\title{An Angular Diameter Measurement of $\beta$ UMa via Stellar Intensity Interferometry with the VERITAS Observatory}

\author[0000-0002-2028-9230]{A.~Acharyya}\affiliation{Department of Physics and Astronomy, University of Alabama, Tuscaloosa, AL 35487, USA}
\author[0000-0002-4104-7580]{J.~P.~Aufdenberg}\affiliation{Embry-Riddle Aeronautical University, Physical Sciences Department, 1 Aerospace Blvd, Daytona Beach, FL 32114, USA}
\author[0000-0002-3886-3739]{P.~Bangale}\affiliation{Department of Physics and Astronomy and the Bartol Research Institute, University of Delaware, Newark, DE 19716, USA}
\author[0000-0002-9675-7328]{J.~T.~Bartkoske}\affiliation{Department of Physics and Astronomy, University of Utah, Salt Lake City, UT 84112, USA}
\author{P.~Batista}\affiliation{DESY, Platanenallee 6, 15738 Zeuthen, Germany}
\author[0000-0003-2098-170X]{W.~Benbow}\affiliation{Center for Astrophysics $|$ Harvard \& Smithsonian, Cambridge, MA 02138, USA}
\author{A.~J.~Chromey}\affiliation{Center for Astrophysics $|$ Harvard \& Smithsonian, Cambridge, MA 02138, USA}
\author{J.~D.~Davis}\affiliation{University of Utah, Department of Physics and Astronomy, 115 South 1400 East 201, Salt Lake City, Utah, USA}
\author[0000-0001-6674-4238]{Q.~Feng}\affiliation{Center for Astrophysics $|$ Harvard \& Smithsonian, Cambridge, MA 02138, USA}
\author[0000-0002-2944-6060]{G.~M.~Foote}\affiliation{Department of Physics and Astronomy and the Bartol Research Institute, University of Delaware, Newark, DE 19716, USA}
\author[0000-0003-1614-1273]{A.~Furniss}\affiliation{Department of Physics, California State University - East Bay, Hayward, CA 94542, USA}
\author[0000-0002-0109-4737]{W.~Hanlon}\affiliation{Center for Astrophysics $|$ Harvard \& Smithsonian, Cambridge, MA 02138, USA}
\author[0000-0001-6951-2299]{C.~E.~Hinrichs}\affiliation{Center for Astrophysics $|$ Harvard \& Smithsonian, Cambridge, MA 02138, USA and Department of Physics and Astronomy, Dartmouth College, 6127 Wilder Laboratory, Hanover, NH 03755 USA}
\author[0000-0002-6833-0474]{J.~Holder}\affiliation{Department of Physics and Astronomy and the Bartol Research Institute, University of Delaware, Newark, DE 19716, USA}
\author[0000-0002-1089-1754]{W.~Jin}\affiliation{Department of Physics and Astronomy, University of California, Los Angeles, CA 90095, USA}
\author[0000-0002-3638-0637]{P.~Kaaret}\affiliation{Department of Physics and Astronomy, University of Iowa, Van Allen Hall, Iowa City, IA 52242, USA}
\author{M.~Kertzman}\affiliation{Department of Physics and Astronomy, DePauw University, Greencastle, IN 46135-0037, USA}
\author[0000-0003-4785-0101]{D.~Kieda}\affiliation{University of Utah, Department of Physics and Astronomy, 115 South 1400 East 201, Salt Lake City, Utah, USA}
\author[0000-0002-4260-9186]{T.~K.~Kleiner}\affiliation{DESY, Platanenallee 6, 15738 Zeuthen, Germany}
\author[0000-0002-4289-7106]{N.~Korzoun}\affiliation{Department of Physics and Astronomy and the Bartol Research Institute, University of Delaware, Newark, DE 19716, USA}
\author[0000-0002-3489-7325]{T.~LeBohec}\affiliation{University of Utah, Department of Physics and Astronomy, 115 South 1400 East 201, Salt Lake City, Utah, USA}
\author[0000-0001-5047-5213]{M.~A.~Lisa}\affiliation{The Ohio State University, Department of Physics, 191 W Woodruff Ave, Columbus, Ohio, USA}
\author[0000-0003-3802-1619]{M.~Lundy}\affiliation{Physics Department, McGill University, Montreal, QC H3A 2T8, Canada}
\author[0000-0002-3687-4661]{N.~Matthews}\affiliation{Universit\'{e} C\^{o}te d'Azur, CNRS, Institut de Physique de France, France} \affiliation{Space Dynamics Laboratory, Utah State University, Logan, UT, USA}
\author{C.~E~McGrath}\affiliation{School of Physics, University College Dublin, Belfield, Dublin 4, Ireland}
\author[0000-0001-7106-8502]{M.~J.~Millard}\affiliation{Department of Physics and Astronomy, University of Iowa, Van Allen Hall, Iowa City, IA 52242, USA}
\author[0000-0002-1499-2667]{P.~Moriarty}\affiliation{School of Natural Sciences, University of Galway, University Road, Galway, H91 TK33, Ireland}
\author{S.~Nikkhah}\affiliation{The Ohio State University, Department of Physics, 191 W Woodruff Ave, Columbus, Ohio, USA}
\author[0000-0002-9296-2981]{S.~O'Brien}\affiliation{Physics Department, McGill University, Montreal, QC H3A 2T8, Canada and Arthur B. McDonald Canadian Astroparticle Physics Research Institute, 64 Bader Lane, Queen's University, Kingston, ON Canada, K7L 3N6}
\author[0000-0002-4837-5253]{R.~A.~Ong}\affiliation{Department of Physics and Astronomy, University of California, Los Angeles, CA 90095, USA}
\author[0000-0001-7861-1707]{M.~Pohl}\affiliation{Institute of Physics and Astronomy, University of Potsdam, 14476 Potsdam-Golm, Germany and DESY, Platanenallee 6, 15738 Zeuthen, Germany}
\author[0000-0002-0529-1973]{E.~Pueschel}\affiliation{DESY, Platanenallee 6, 15738 Zeuthen, Germany}
\author[0000-0002-4855-2694]{J.~Quinn}\affiliation{School of Physics, University College Dublin, Belfield, Dublin 4, Ireland}
\author{P.~L.~Rabinowitz}\affiliation{Department of Physics, Washington University, St. Louis, MO 63130, USA}
\author[0000-0002-5351-3323]{K.~Ragan}\affiliation{Physics Department, McGill University, Montreal, QC H3A 2T8, Canada}
\author{E.~Roache}\affiliation{Center for Astrophysics $|$ Harvard \& Smithsonian, Cambridge, MA 02138, USA}
\author[0009-0000-0003-6407]{J.~G.~Rose}\affiliation{The Ohio State University, Department of Physics, 191 W Woodruff Ave, Columbus, Ohio, USA}
\author[0009-0006-8589-6775]{J.~L.~Sackrider}\affiliation{Embry-Riddle Aeronautical University, Physical Sciences Department, 1 Aerospace Blvd, Daytona Beach, FL 32114, USA}
\author[0000-0003-1387-8915]{I.~Sadeh}\affiliation{DESY, Platanenallee 6, 15738 Zeuthen, Germany}
\author[0000-0002-3171-5039]{L.~Saha}\affiliation{Center for Astrophysics $|$ Harvard \& Smithsonian, Cambridge, MA 02138, USA}
\author{G.~H.~Sembroski}\affiliation{Department of Physics and Astronomy, Purdue University, West Lafayette, IN 47907, USA}
\author[0000-0002-9856-989X]{R.~Shang}\affiliation{Department of Physics and Astronomy, Barnard College, Columbia University, NY 10027, USA}
\author{D.~Tak}\affiliation{DESY, Platanenallee 6, 15738 Zeuthen, Germany}
\author[0009-0004-7242-0204]{M.~Ticoras}\affiliation{The Ohio State University, Department of Physics, 191 W Woodruff Ave, Columbus, Ohio, USA}
\author{J.~V.~Tucci}\affiliation{Department of Physics, Indiana University-Purdue University Indianapolis, Indianapolis, IN 46202, USA}
\author[0000-0002-2730-2733]{S.~L.~Wong}\affiliation{Physics Department, McGill University, Montreal, QC H3A 2T8, Canada}

\collaboration{60}{The VERITAS Collaboration}

\begin{abstract}
\nolinenumbers
We use the VERITAS imaging air Cherenkov Telescope (IACT) array to obtain the first measured angular diameter of $\beta$ UMa at visual wavelengths using stellar intensity interferometry (SII) and independently constrain the limb-darkened angular diameter. 
The age of the Ursa Major moving group has been assessed from the ages of its members, including nuclear member Merak ($\beta$ UMa), an A1-type subgiant, by comparing effective temperature and luminosity constraints to model stellar evolution tracks.  
Previous interferometric limb-darkened angular-diameter measurements of $\beta$ UMa in the near-infrared (CHARA Array, 1.149 $\pm$ 0.014\,mas) and mid-infrared (Keck Nuller, 1.08 $\pm$ 0.07\,mas), together with the measured parallax and bolometric flux, have constrained the effective temperature.
This paper presents current VERITAS-SII observation and analysis procedures to derive squared visibilities from correlation functions.
We fit the resulting squared visibilities to find a limb-darkened angular diameter of $1.07 \pm 0.04 {\rm ~(stat)} \pm 0.05$~(sys)~mas, using synthetic visibilities from a stellar atmosphere model that provides a good match to the spectrum of $\beta$ UMa in the optical wave band.  The VERITAS-SII limb-darkened angular diameter yields an effective temperature of $9700\pm200\pm 200$ K,
consistent with ultraviolet spectrophotometry, and an age of $390\pm 29 \pm 32 $ Myr, using MESA Isochrones and Stellar Tracks (MIST). This age is consistent with 408 $\pm$ 6 Myr from the  CHARA Array angular diameter. 
\end{abstract}

\correspondingauthor{Mackenzie Ticoras}\email{scott.2275@buckeyemail.osu.edu}

\keywords{Long baseline interferometry (932), Fundamental parameters of stars (555), Astronomy data modeling (1859)}

\section{Introduction} \label{sec:Intro}

Stellar kinematic groups are an evolutionary link between clusters and field stars. 
Age estimates for such groups are not only of interest with regards to stellar evolution but also planetary evolution. 
Observations of exoplanets orbiting a member of the Ursa Major moving group help to constrain planet evolution during the first billion years \citep{Mann_2020}.  
As a nuclear member of the Ursa Major moving group \citep{soderblom1993,king_2003}, $\beta$ UMa (\object{Merak}, HD 95418, HR 4295)  was used, along with six other A-type stars, to constrain the age of the group to 414 $\pm$ 23 Myr by \cite{Jones_2015}. 
$\beta$ UMa stands apart from the other six A-type stars in being an apparently slow rotator ($v\sin i = 46\,\rm{km\, s^{-1}}$). 
High-resolution spectra suggest $\beta$ UMa is unlikely to be viewed nearly pole-on as gravity darkening would produce peculiar spectral line profiles as seen in the case of the nearly pole-on Vega \citep{Hill2010}. 
As a result, it may be possible to more tightly constrain $\beta$ UMa's fundamental parameters relative to the faster-spinning nuclear group members.

Michelson interferometers at both the Center for High Angular Resolution Astronomy (CHARA) array and the Keck Observatory have measured the angular diameter of $\beta$ UMa \citep{Boyajian_2012, Mennesson_2014}, see Table \ref{tab:Literature_Angular_Diameters}.
The CHARA CLASSIC beam combiner measurement at 2.141 $\pm$ 0.003\,$\mu$m yielded a limb-darkened angular diameter, $\theta_{\rm LD}$, of 1.149 $\pm$ 0.014\,mas, while the Keck Interferometer mid-infrared Nulling (KIN) instrument measurements, in 10 spectral channels covering the 8 - 13\,$\mu$m range, yielded $\theta_{\rm LD}$ = 1.08 $\pm$ 0.07\,mas.
The mid-infrared KIN measurement should be insensitive to limb darkening, the Rayleigh-Jeans law indicates a change in spectral radiance with temperature is proportional to $\lambda^{-4}$, and the limb-darkening correction for the CHARA Array measurement is less than 2\%. 
The uncertainty in the angular diameter dominates the uncertainty on $\beta$ UMa's effective temperature and radius as the bolometric flux from \cite{Boyajian_2012} is uncertain at the 2\% level and the parallax uncertainty is less than 1\% \citep{V07}.  
Below, we describe an independent measurement of the angular diameter of $\beta$ UMa using intensity interferometry, the first angular-diameter measurement of the star at visible wavelengths. 
This star was not observable from the Intensity Interferometer at Narrabri, Australia \citep{1974RHB_book} due to its declination ($\delta = +56.3$). 

\begin{deluxetable*}{cccc}[h]
\label{tab:Literature_Angular_Diameters}
\tablecaption{Angular diameter measurements and estimates for $\beta$ UMa}
\tablecolumns{4}
\tablewidth{0pt}
\tablehead{
\colhead{Reference/Method} &
\colhead{Wavelength ($\mu$m)} &
\colhead{$\theta_{\rm UD}$ (mas)} & 
\colhead{$\theta_{\rm LD}$ (mas)}}
\startdata
 \multicolumn{4}{c}{Interferometric measurements}\\
 \hline
VSII (This work)   & 0.416  & 1.01 $\pm$ 0.03 $\pm$ 0.05 & 1.07 $\pm$ 0.04 $\pm$ 0.05 \\ 
CHARA Array \citep{Boyajian_2012} & 2.14  & 1.133 $\pm$ 0.014 & 1.149 $\pm$ 0.014\\
Keck Interferometer Nuller \citep{Mennesson_2014} & 8 - 13  &1.078 $\pm$ 0.065  & 1.078 $\pm$ 0.065 \\
\hline
 \multicolumn{4}{c}{Indirect estimates}\\
\hline
Bolometric flux - T$_{\rm eff}$ relation \citep{Zorec2009} &0.1380 - 1.108  &\nodata &1.112 $\pm$ 0.007\\
V vs. V-K relation \citep{JMMC2010} &0.442  & 0.97 $\pm$ 0.07 & 1.02 $\pm$ 0.07\\ 
Spectral energy distribution \citep{Swihart2017} &$0.2-10$ &\nodata  &1.050 $\pm$ 0.063 \\
\enddata
\tablecomments{$\theta_{\rm UD}$ is the uniform disk angular diameter at the given
wavelength/waveband.  $\theta_{\rm LD}$ is the limb-darkened angular diameter. Limb darkening is expected to be negligible at 8 - 13 $\mu$m.} 
\end{deluxetable*}

The paper is organized as follows. 
Section \ref{sec:datades} provides a description of the VERITAS Stellar Intensity Interferometer (VSII). 
The VSII observations of $\beta$ UMa are described in Section \ref{sec:obs}. 
Section \ref{sec:DataAnalysis} outlines the analyses to extract the source angular size.
The angular size estimate is folded into stellar calculations to estimate the age of $\beta$ UMa in Section~\ref{sec:Age}.
Finally, Section \ref{sec:discussion} discusses the implications of the results and future plans for VSII.
An independent secondary analysis of the stellar diameter, based on a likelihood approach, is presented in the appendix; it confirms the results of the primary analysis discussed here.

\section{The VERITAS Stellar Intensity Interferometer \label{sec:datades}}

The Very Energetic Radiation Imaging Telescope Array System (VERITAS) is composed of four 12-m diameter Imaging Air Cherenkov Telescopes (IACTs) used for very high energy ($E>100\,\rm GeV$) gamma-ray astronomy~\citep{HOLDER2006391}. 
The four telescopes are individually referred to as T1, T2, T3, and T4. During bright moonlight periods, when normal gamma-ray observations are extremely restricted, the array is used for stellar intensity interferometry (SII) observations. 
For this, an SII instrumentation plate is mounted in the focal plane of each telescope. The SII plate is designed to perform spectral filtering and high-time-resolution detection of the starlight intensity. 
The starlight is passed through a narrowband interferometric filter with a central transmitted wavelength of 416\,nm and an optical bandpass of $\sim13\,\rm nm$. 
The filtered light is focused on a photo-multiplier tube (PMT), and the resulting signal is sent via co-axial cable after amplification to a high-speed data acquisition system located in the counting house adjacent to each telescope.
During observation, the analog signal is digitized continuously with a sampling time of 4\,ns, and the data is streamed to disk at each telescope independently in a compressed file format. All digitizer sampling clocks are phase synchronized to a central White Rabbit 10 MHz clock that is distributed from the main operation building via optical fibers. 
Collectively, the four telescopes generate approximately 3.5 terabytes of raw data for every hour of observation. 
After observations are completed, the normalized correlation function for each of up to six telescope pairs is computed. 
Correlation functions are obtained either using Field-Programmable Gate Arrays (FPGAs) or via software on multiple CPU cores.  
For the same raw data, the two systems produce correlation functions that are identical to one another. 

A peak in the correlation function reveals the second-order degree of spatial coherence between the light collected by the two telescopes. 
For thermal light, the second-order degree of coherence is related to the first-order degree of coherence by the Siegert relation~\citep{Siegert1943}. The first-order degree of coherence can be recognized as the interferometric visibility that is related to the angular brightness distribution of the target by the van Cittert-Zernike theorem.
Inversely, measurements of the squared visibility from the telescope pairwise correlation functions can be analyzed to characterize the brightness distributions over angular scales given by the ratio between the wavelength and the inter-telescope baselines.   

The normalized correlation functions are computed with the following: 
\begin{equation} 
        g^{2}_{AB}(\tau) = \frac{\langle I_{A}(t) I_{B}(t+\tau) \rangle}{\langle I_{A}(t) \rangle \langle I_{B}(t) \rangle},
        \label{eq:IntensityCorrelation}
\end{equation}
where $I_{A/B}$ is the intensity measured at telescope $A$ or $B$, $\tau$ is the relative time-delay between signals $A$ and $B$, and the brackets indicate an average over a time interval $T_{1/8}=0.125\,\rm s$. 
For every time $T_{1/8}$ of data and for each telescope pair, a correlation function over 128 relative time delay bins, each of width 4\,ns, is stored and archived alongside time stamps and the average photo-current in each telescope as the final data product to be later analyzed. 

The relative time lag $\tau_0$, for which a correlation peak is expected, is determined by two factors.
The first is the optical path difference of starlight reaching the two telescopes; this is straightforward to account for, as we discuss in Section~\ref{subsec:OPDShift}. 
The second is the difference between the start times of the independent data acquisition systems of the telescopes.
In the current system, this start-time difference can vary by as much as 20\,ns from one observation ``run" to the next.
As discussed in Section~\ref{sec:QuantifyingVisibility}, this leads to difficulties when measuring very weak signals (e.g. at large baselines).

\begin{deluxetable*}{cccc}
\label{tab:observations}
\tablecaption{VERITAS SII Observations of Merak
}
\tablecolumns{4}
\tablewidth{0pt}
\tablehead{
\colhead{UTC Date} &
\colhead{Telescope Pairs Used} &
\colhead{Observation Time} &
\colhead{Typical Moon Angle}\\
\colhead{(dd/mm/yyyy)} &
\colhead{($T_A$,$T_B$)}& 
\colhead{(hours)} &
\colhead{(degrees)}}
\startdata
  17/12/2021 & (1,2), (2,3), (2,4), (3,4)  & 8.3 & 78 \\
  12/02/2022 & (1,2), (2,3), (2,4), (3,4)  & 11.7 & 59 \\
  14/02/2022 & (1,2), (2,3), (2,4), (3,4)  & 11.2 & 45 \\
  13/03/2022 & (1,2), (2,4), (3,4)         & 6.2 & 47 \\
\enddata
\tablecomments{Observations times are before data quality cuts on the parameters of correlation peak have been applied, see Section \ref{sec:QuantifyingVisibility}. 
}
\end{deluxetable*}

\section{Observations and data acquired}\label{sec:obs}

Data were accumulated during five nights between December 2021 and March 2022, as detailed in Table \ref{tab:observations}. 
The weather on all nights was clear with only very slight wispy cloud cover on December $\rm 17^{th}$ and March $\rm 13^{th}$. 
Figure~\ref{fig:uvcover} shows the baseline coverage in the Fourier plane for these observations.
  
\begin{figure*}[h!]
  \begin{center}
  \includegraphics[width=0.8\linewidth]{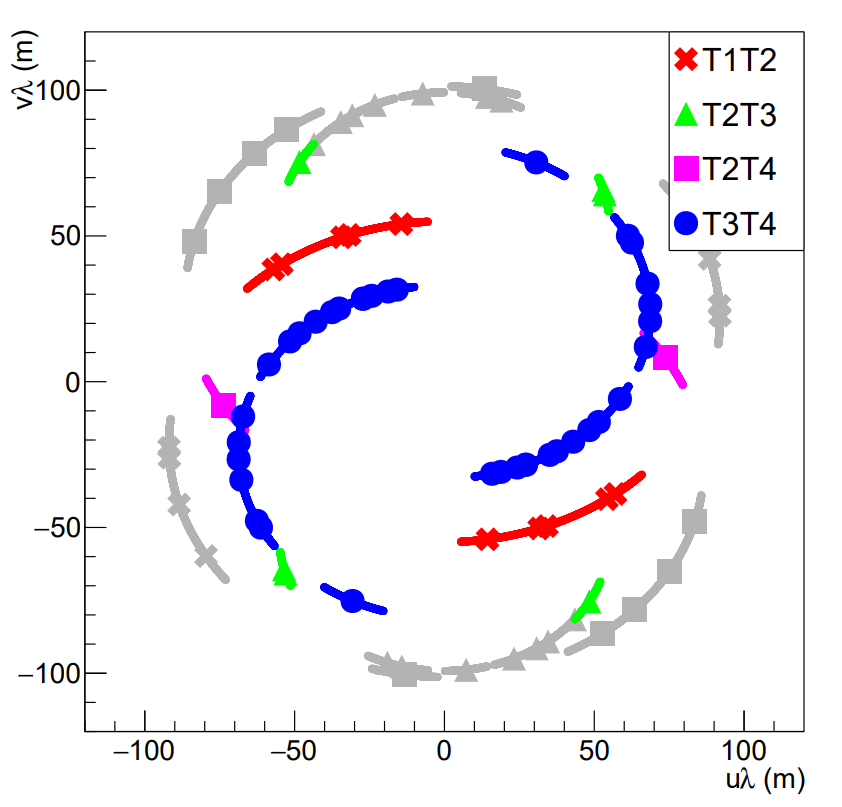}
  \end{center}
  \caption{Telescope pair baseline coverage in the Fourier plane for the data points in Figure~\ref{fig:UniformDiskVisibilites}.
  Points indicate the baseline halfway into the run; lines indicate the range of baselines covered during a run, and gray tracks represent baselines near 100 m, treated as a single point with zero area in Figure~\ref{fig:UniformDiskVisibilites}.} 
  \label{fig:uvcover}
\end{figure*}

Intensity interferometry requires bright sources in order to allow for a  measurement of spatial coherence, and smaller sources produce strong signals over a larger range of baselines. 
For the source studied here ($\beta$~UMa), telescope pairs with baselines $\gtrsim100$\,m do not produce correlation peaks large enough to emerge above noise.
Because of the run-to-run variation in $\tau_0$ discussed in Section~\ref{sec:datades} (c.f. Figure~\ref{fig:HBTPeaks}), in the absence of a clearly identifiable correlation peak, it is impossible to make a meaningful estimate of the squared visibility on a run-by-run basis.
In Section~\ref{sec:QuantifyingVisibility}, we discuss cuts used to identify good correlation peaks, and how we treat long-baseline runs.

Some stray light contributes to the digitized PMT current
and reduces the observed strength of the correlation.
To obtain a measurement of the second-order degree of coherence corresponding only to starlight, a correction (c.f. Section~\ref{subsec:BackgroundIntensity}) taking into account the night-sky background is applied to the correlation functions. 
To estimate the background light not associated with the star, short duration (one minute) ``off'' runs are taken every 1-2 hours, with the telescopes pointing $0.5^\circ$ away from the target.  
Typically, the intensity recorded is $\sim3\%$ of that observed when the telescope is pointed at the star, and the effect on the extracted stellar radius is very small.

During the 0.5 to 2-hour duration of the on-target runs, the telescope pairs' projected baselines vary significantly as the target transits the sky as shown in Figure~\ref{fig:uvcover}.
The optical path delay discussed in Section~\ref{sec:datades} also changes during the observation.
This is evident in the series of correlation functions for a two-hour observation with telescopes 3 and 4 shown in Figure~\ref{fig:HeatMapOSU}. 
The evolution of the optical path delay is accounted for in the analysis discussed below; c.f. section~\ref{subsec:OPDShift}.

\begin{figure*}[ht!]
    \includegraphics[width=\linewidth]{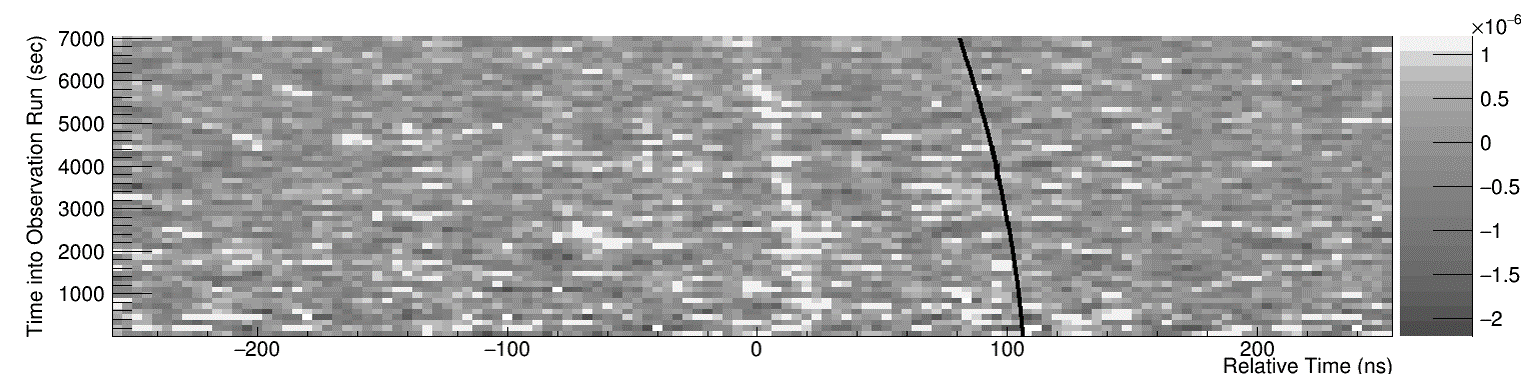} 
    \caption{A series of correlation functions taken during a two-hour run for telescopes 3 and 4.
    Plotted is $g^{2}_{3,4}(\tau,t)-1$, the correlation function defined in Equation~\ref{eq:DarkRunFactor} as function of time lag ($\tau$) and time-in-run ($t$) on the horizontal and vertical axes, respectively.
    The position in relative time of the small ($\sim10^{-6}$) but significant correlation peak varies during the run, as the geometric optical path delay (OPD) varies.
    A calculated OPD, based on knowledge of the star's position, is indicated with a black curve, which has been offset from the data peak for clarity. }
  \label{fig:HeatMapOSU}
\end{figure*}

Some runs feature apparent ``radio-frequency" contamination of the correlation functions.  
This generally consists of multiple narrow-band signals of differing frequencies. 
In particular, a signal of 79.5\,MHz frequency occasionally dominates the correlation function.
Detailed study indicates that this signal originates from a single location on the VERITAS site, likely the microwave network link, and the ``radio-frequency" noise is a 79.5\,MHz burst pattern of $\sim10$~GHz radiation. 
This beacon has well-defined characteristics and the contamination signal changes as a function of the telescope pointing direction. 
There are other less significant contaminating frequencies coming from instrumentation within the telescopes. These signals are removed from the data during analysis, as discussed in Section~\ref{subsec:NoiseFrequencies}.

\section{Analysis}\label{sec:DataAnalysis}
 
The following subsections present the major steps of the analysis to obtain measurements of the squared visibilities as a function of the projected baseline between pairs of telescopes. 
The squared visibilities are then fitted to extract a uniform disk angular diameter for this wavelength and the limb-darkened angular diameter. 

\subsection{Correction of effects from stray light and dark currents} \label{subsec:BackgroundIntensity}

When tracking $\beta$~UMa, the recorded currents from the PMTs are dominated by the light from the star. 
However, the signals also contain a small contribution from (e.g.) moonlight scattered by the atmosphere and instrumental noise.
The ``off" runs (see Section~\ref{sec:obs}) are used to quantify and correct for this.

The stray contribution may vary over the course of a run.
The night-sky current in telescope $A$ at a given time, $ I_{A}^{\rm off}(t)$, is estimated by linearly interpolating between the currents measured in ``off" runs immediately before and after the on-target run.
The correlation function due only to the starlight is then calculated according to
\begin{equation}
    \label{eq:DarkRunFactor}
    g^{2}_{AB}(\tau,t)-1 = \left(g^{(2*)}_{AB}(\tau,t)-1\right)
    \cdot\frac{\langle I_A(t)\rangle\langle I_B(t)\rangle}{\left(\langle I_A(t)\rangle-\langle I^{\rm off}_A(t)\rangle\right)\left(\langle I_B(t)\rangle-\langle I^{\rm off}_B(t)\rangle\right)}, 
\end{equation}
where $g^{(2*)}_{AB}(\tau,t)$ represents the correlation function before the ``off" runs are removed and $g^{(2)}_{AB}(\tau,t)$ is the corrected result. 
Overall, the difference between neglecting or accounting for the extraneous contribution is about 2\% for the stellar diameter estimates discussed below.

\subsection{Removal of extraneous frequency contamination} \label{subsec:NoiseFrequencies}

As discussed in Section~\ref{sec:obs}, the correlation functions from some runs include ``radio" contamination at MHz frequencies.
To remove them, a Fourier transform was performed on the digitized currents to generate a catalog of contaminating frequencies, $\left\{\omega_i\right\}$.

Then, Fourier series are used to establish the amplitudes of the cosine and sine components for each radio frequency in every 32\,s time slice of the data, per 
\begin{equation}
\label{eq:CosineSine}
    A_C(\omega_i,t) \equiv \frac{\sum_j \cos(\omega_i\tau_j) g_{AB}^{2}(\tau_j,t)}{\sum_j \cos^2(\omega_i\tau_j)},
    \qquad\qquad
    A_S(\omega_i,t) \equiv \frac{\sum_j \sin(\omega_i\tau_j) g_{AB}^{2}(\tau_j,t)}{\sum_j \sin^2(\omega_i\tau_j)},
\end{equation}
where the sum is over relative time delay bins indexed by $j$. 
Importantly, these sums in Equation~\ref{eq:CosineSine} are restricted to the values of the discrete relative time delays $\tau_j$ outside the region where the coherence peak can be located. 
This is to ensure the correlation peak does not bias the characterization of the radio contamination. These moments are then combined into a single function,
\begin{equation} 
g^{(2,{\rm Radio})}_{AB}(\tau,t) = \sum_{i}
\sqrt{A_C^2(\omega_i,t)+A_S^2(\omega_i,t)}\cos\left(\omega_i\tau - \tan^{-1} \left(\frac{A_S(\omega_i,t)}{A_C(\omega_i,t)}\right)\right),
\end{equation}
which is then subtracted from $g^{2}_{AB}(\tau,t)$.
Propagation of statistical uncertainties on data points outside the peak region leads to $\sim4\%$ increase of uncertainties on data points near the peak. 

\subsection{Optical path delay shifting and time-averaging} \label{subsec:OPDShift}

As discussed in Section~\ref{sec:obs} and shown in Figure~\ref{fig:HeatMapOSU}, the telescope positions and local sky coordinates of the star determine the relative time delay shift of the coherence peak as a function of time, $\Delta_{AB}(t)$.
A time-averaged correlation function for a run is obtained by averaging the relative time delay shifted correlation functions resulting from $\tfrac{1}{8}$-second time slices. The average over $N$ time slices is obtained as 
\begin{equation}
\label{eq:TimeAverageG2}
    \langle g^{2}_{AB}(\tau)\rangle=\frac{1}{N}\sum_{i=1}^N g_{AB}^{2}(\tau-\Delta(t_i),t_i).
\end{equation}

Because the time shifts ($\Delta$) vary by less than 1~ns from one $T_{1/8}$ time slice to the next, the time-averaged correlation function defined in Equation~\ref{eq:TimeAverageG2} may be binned more finely in relative time delay $\tau$ than the 4~ns sampling time of the data acquisition system.
Since the temporal width of the coherence peak (determined by PMT/pre-amp bandwidth and mirror optics) is also about 4~ns~\citep{natureSIIDemo}, binning more finely allows better resolution of the peak.
Time-averaged correlation functions, with 2~ns bins in $\tau$, for two runs with telescopes 3 and 4, are shown in Figure~\ref{fig:HBTPeaks}.

\begin{figure*}[h!]
  \begin{minipage}[b]{0.48\textwidth}
    \includegraphics[width=\textwidth]{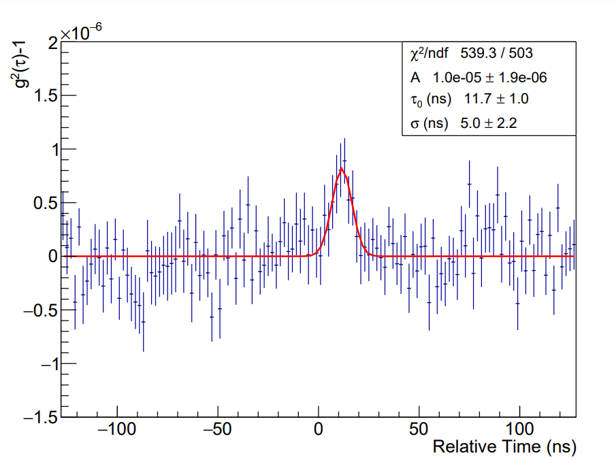}
  \end{minipage}
  \hfill
  \begin{minipage}[b]{0.48\textwidth}
    \includegraphics[width=\textwidth]{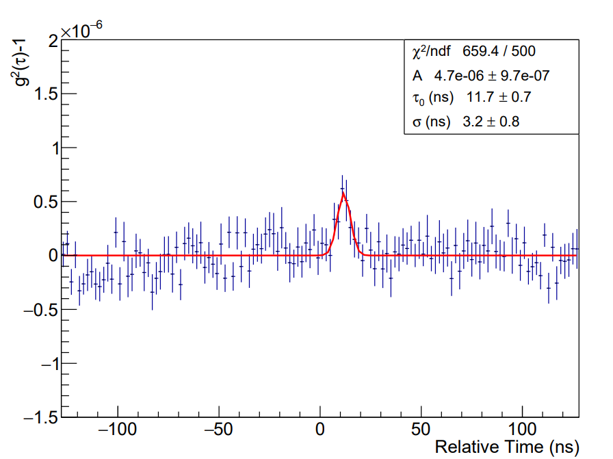}
  \end{minipage}
  \caption{Time-averaged correlation functions $\langle g^{2}_{34}(\tau)\rangle$ from telescopes 3 and 4.
   Data from the left (right) panel was taken on December $16^{\rm th}$, 2021 (February $11^{\rm th}$, 2022).
   The projected baseline for the later run was larger than that for the earlier run, leading to a smaller correlation peak.  
   The peaks are fit with a Gaussian functional form to quantify the correlation strength.}
   \label{fig:HBTPeaks}
\end{figure*}

\newpage
\subsection{Quantifying the squared visibility}
\label{sec:QuantifyingVisibility}

\begin{figure*}[t!]  
  \centering{\includegraphics[width=0.8\linewidth]{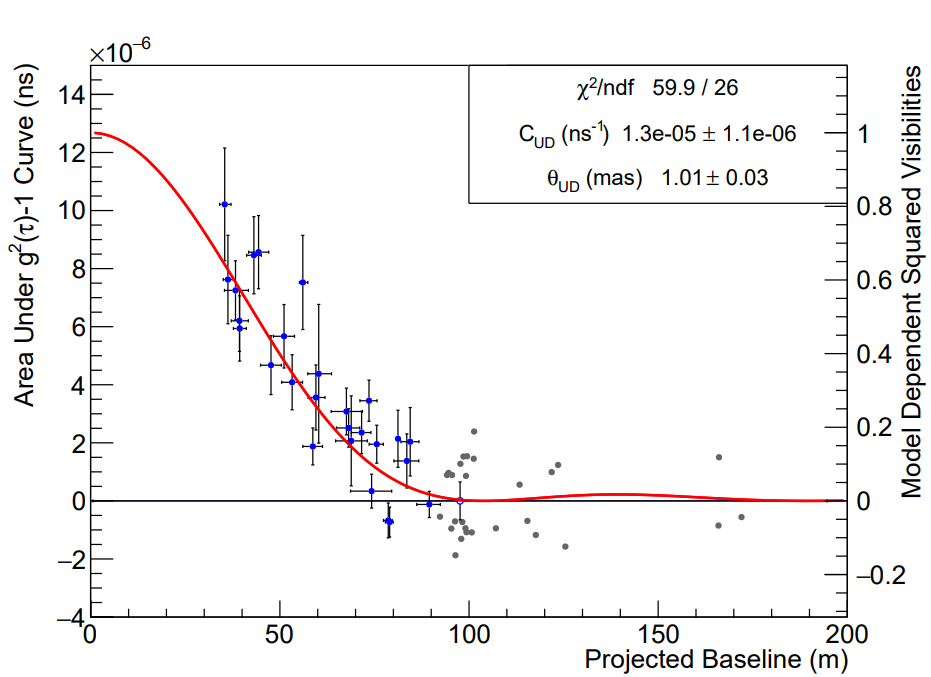}}
  \caption{The measurements of the squared visibilities obtained as described in Subsection \ref{sec:QuantifyingVisibility} are presented as a function of the corresponding projected baseline lengths for all pairs of telescopes. The red curve represents the fit of the uniform disk model to the selected data points indicated in blue color (see text for details). The dark grey points do not directly contribute to the red fit curve and their contribution is discussed in Subsection \ref{sec:QuantifyingVisibility}.}
  
  \label{fig:UniformDiskVisibilites}
\end{figure*}

The integral of the coherence peak in the time-averaged correlation function is proportional to the squared visibility.
Because the location of the correlation peak may vary by as much as $\pm$ 10\,ns from run to run (see Section~\ref{sec:datades}), we use the Minuit package~\citep{James:1975dr} to fit a Gaussian functional form within a 20\,ns-wide range in $\tau$.  
The Gaussian width is expected to be about 3.5 - 4\,ns and is allowed to vary between 2.5\,ns and 5.5\,ns.  
The amplitude is unconstrained, both in magnitude and in sign, to avoid biasing the fitting of fluctuations when a strong enough coherence peak is not present.

Large peaks, such as that in the left panel of Figure~\ref{fig:HBTPeaks}, are easily found.
However, the inability to tightly constrain the peak position is a significant problem for weaker coherence peaks, as the peak competes with fluctuations over a 20\,ns-wide window.

To avoid human bias and keep the analysis simple, we apply the following peak quality cuts.
After $\langle g^{2}_{AB}(\tau)\rangle$ is fitted, if the Gaussian width is not in the range (2.5\,ns -- 5.5\,ns),
and/or if the peak center is not within $\pm10$~ns of the nominally expected position, the result is discarded.
All filled data points (blue and grey) in Figure~\ref{fig:UniformDiskVisibilites} have passed these simple quality cuts.
The integral of the fitted Gaussian is used as a proxy for the squared visibility, $|V|^2$, and plotted versus the average projected baseline in Figure~\ref{fig:UniformDiskVisibilites}.
Vertical error bars represent the statistical uncertainties on the fitted coherence peak integral, while horizontal error bars denote the range of baselines for a given run/telescope pair.
The grey points, which are drawn without error bars for clarity, require further discussion.

\begin{figure}[t!]  
  \centering{\includegraphics[width=0.7\linewidth]{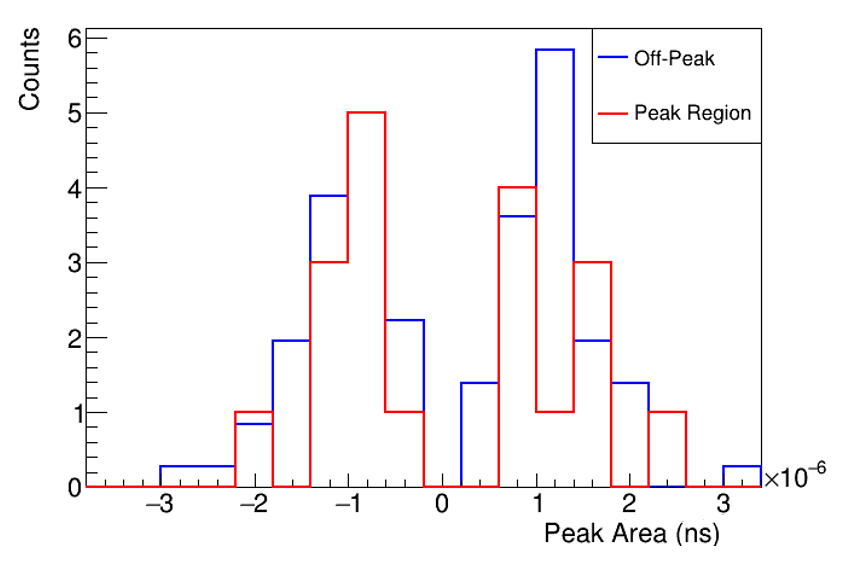}}
  \caption{Red: The distribution of peak amplitudes returned by the peak-finding/fitting algorithm corresponding to the grey data-points on Figure~\ref{fig:UniformDiskVisibilites} for baselines between 90-105~m. 
   As with all data shown in Figure~\ref{fig:UniformDiskVisibilites},
   the peak centers were within $\pm$10\,ns of nominal expectation.
   Blue: the same distribution, for the same runs, but using regions well {\it outside} the relative time delay $\tau$ region of any correlation signal.
  All peaks included here passed the width quality cut described in the text.
  A Kolmogorov-Smirnov test estimates the probability that the two distributions are consistent with the same parent is 99\% 
  The off-peak distribution is renormalized to have the same area as the red curve in this figure, for proper comparison.
  \label{fig:Muck}
}
\end{figure}

If the position of the coherence peak in time lag ($\tau$) were well-known, then a correlation of arbitrarily small magnitude could be measured, though it may be zero within well-defined uncertainties.
However, in our measurement, the centroid of the expected peak is not well constrained for a given run/pair.
Hence, the peak-finding/fitting algorithm may converge on fluctuations, resulting in a found peak that passes our quality cuts.
In fact, for a zero-amplitude correlation (i.e. the absence of any true correlation), our algorithm will find peaks of positive or negative amplitude with equal probability and will tend {\it not} to report a peak of zero magnitude, even though zero is the correct value.

Figure~\ref{fig:Muck} illustrates this quantitatively.
In red is the histogram of the values of peak amplitude from the grey data points on Figure~\ref{fig:UniformDiskVisibilites} with baselines between 90\,m and 105\,m.
The blue histogram shows the same distribution, but now running the algorithm on regions of the correlation functions well outside the $\pm10$\,ns window where any coherence peak lies. 

Thus, the blue histogram represents the peak amplitude distribution in the absence of coherence. 
This distribution is bimodal as a consequence of the fitting procedure 
biasing the peak amplitude away from zero in the absence of a strong enough coherence peak. 
This makes direct inclusion of each low squared visibility point into a model fit inappropriate.  However, the similarity between the two distributions (a Kolmogorov-Smirnov test reveals a 99\% 
probability that these distributions arise from the same parent) leads us to conclude that the strength of the correlation at a baseline $\sim100\rm\,m$ is consistent with zero.
Hence, when doing the data-model fitting, we replace the cluster of points with projected baselines between 90\,m and 105\,m with a single datapoint at $b=98$\,m 
and ${\rm area}=0$\,ns; where the error bar is estimated from simulation studies and determined to be $0.6\times10^{-6}$\,ns.
This data point, shown as an open blue symbol in Figure~\ref{fig:UniformDiskVisibilites}, is included together with the filled blue data points when performing model fits discussed below.

The same reasoning would almost certainly lead to the same conclusion for the data-point clusters around baselines $\sim125$\,m and at $\sim170$\,m in Figure~\ref{fig:UniformDiskVisibilites}.
However, our treatment of the 90\,m - 105\,m data point cluster resulted in squared visibility consistent with zero, based on a {\it distribution} of peak amplitude in the absence of coherence. 
In the case of the $\sim125$\,m and $\sim170$\,m data point clusters, there are too few points to say anything meaningful. 
Consequently, these points are ignored when extracting the source size below.

\subsection{Extracting stellar angular diameters}
\label{sec:ExtractingDiameters}

The squared visibility as a function of the baselines corresponds to the squared magnitude of the Fourier transform of the brightness angular distribution of the star. 
In the case of a circularly symmetric target, the squared visibility measurements for different projected baseline lengths $b$ can be used to achieve a measurement of the stellar diameter, $\theta_D$.
Within the context of a model, we infer the angular diameter by comparing (or fitting) modeled squared visibilities $M(\theta_D,b)$ to the measured ones.

During a run, the projected baseline varies continuously, as the star transits the sky.
Hence, each squared visibility measurement (i.e. each data point on Figure~\ref{fig:UniformDiskVisibilites} and Table~\ref{tab:SquaredVisibilities}) is an average over the baselines in the course of the run.
Furthermore, the starlight flux at the telescopes (measured by PMT currents $I_{A}(t)$ and $I_{B}(t)$) varies during the run due to atmospheric effects and also to small telescope tracking inaccuracies.
The high-current periods will have larger statistical weight than the lower-flux periods, and so our measurements are a {\it weighted} average.
For an apples-to-apples comparison, we calculate the same average for the models.
In particular, for each run/telescope pair combination $i$ (i.e. each blue data point in Figure~\ref{fig:UniformDiskVisibilites}), we calculate the mean squared visibility predicted by the model:
\begin{equation}
\label{eq:AveragingModels}
    M(\theta_D)_i=\frac{\int dt\,I_{A,i}(t)\,I_{B,i}(t)\,M(\theta_D,b_i(t))}{\int dt\,I_{A,i}(t)\,I_{B,i}(t)} ,
\end{equation}
where the integral is over the duration of the run.
Depending on the model, the visibility may depend on one or more parameters in addition to $\theta_D$; we suppress these in the formulae, for clarity and generality.  
The best model parameters are determined by minimizing
\begin{equation}
\label{eq:chi2}
    \chi^2 \equiv \sum_i\left(\frac{M(\theta_D)_i - V_i^2}{\sigma_i}\right)^2 ,
\end{equation}
where $V_i^2$ and $\sigma_i$ are the measured squared visibilities and uncertainties for the data points from Figure~\ref{fig:UniformDiskVisibilites}.

Depending on the smoothness of the predicted visibility and the degree to which the intensity and baseline vary over the course of a run, the weighted average of Equation~\ref{eq:AveragingModels} may be simply replaced by the much simpler approximation
\begin{equation}
    M(\theta_D)_i \approx M(\theta_D,\langle b_i\rangle) .
\end{equation}
In all the fits we discuss below, we have verified that the difference in fit parameters between using the exact expression and the approximation is at least 3 orders of magnitude smaller than the statistical uncertainty.
Hence, smooth models may be safely compared to the data points in Figure~\ref{fig:UniformDiskVisibilites} without the need to know the fine internal structure of the VSII runs.

\subsubsection{Diameter of $\beta$ UMa - Uniform disk model}
\label{sec:UniformDiskDiameterPrimary}

The squared visibility for a projected baseline $b$ for a uniformly illuminated disk of diameter $\theta_{\rm UD}$ is
\begin{equation}
\label{eq:UniformDiskModel}
    V_{\rm UD}^2(\theta_{\rm UD},b) = C_{\rm UD}\times\left(\frac{2J_1(\pi b\,\theta_{\rm UD}/\lambda)}{\pi b\, \theta_{\rm UD}/\lambda}\right)^2 ,
\end{equation}
where $\lambda=416$~nm and $J_1$ denotes a bessel function. 
$C_{\rm UD}$ is the proportionality constant between actual squared visibilities and peak integrals measured from correlation functions, which we treat as a fit parameter.
Setting $M=V_{\rm UD}^2$ in Equations~\ref{eq:AveragingModels} and~\ref{eq:chi2}, we find the stellar angular diameter for the uniform disk model to be $\theta_{\rm UD}=1.01\pm0.03$\,mas.
The best-fit model is shown as a red curve in Figure~\ref{fig:UniformDiskVisibilites}.

As discussed extensively above, the uncertainty in correlation peak position inherent in our present system leads to an analysis that depends on quality cuts for the peaks.
There is then a systematic uncertainty associated with these cuts.
The relative time window in which the Gaussian fit was varied between $\pm$ 10\,ns and $\pm$ 16\,ns, which had a less than 1\% effect on the angular diameter measurement.
Additionally, the allowed peak width, which has a central value of 4.5\,ns and a possible range of $\pm$ 1.5\,ns, was varied between $\pm$ 1.5\,ns and $\pm$ 3\,ns. The contribution to the systematic uncertainty from fluctuations of the stray light background ($I^{\rm off}_A(t)$ and $I^{\rm off}_B(t)$, Equation \ref{eq:DarkRunFactor}) was also considered.  Combined together, these systematics have no more than a 5\% effect on the angular diameter measurement. 

\subsubsection{Diameter of $\beta$ UMa - Model with limb darkening}
\label{sec:LimbDarkeningDiameter}

Limb darkening must be accounted for to properly compare stellar angular diameters measured at different wavelengths.
To account for limb-darkening, a series of synthetic visibilities at different limb-darkened diameters were generated for comparison to the VERITAS measurements.
The synthetic visibilities were calculated for each of the mean baselines in Table \ref{tab:SquaredVisibilities} from a weighted mean of visibilities at 276 different wavelengths across the VSII bandpass employing center-to-limb intensity profiles from the PHOENIX stellar atmosphere code \citep{H99} and the computational methods in \cite{Vega06}. 
A spectrum from this model provides a close match to the observed spectrum.
Figure \ref{fig:spectrum_bandpass} shows the measured VSII bandpass together with archival and model spectra.  
The model spectra were matched to the archival spectrum by first converting vacuum wavelengths to air wavelengths, then Doppler shifting the model spectra to account for both the Earth's barycentric radial velocity in the direction of $\beta$ UMa ($-$10.4 $\rm km\, s^{-1}$), on the date of observation, and $\beta$ UMa's heliocentric radial velocity \citep[$-$13.1 $\rm km\, s^{-1}$]{Measured_RV}. 
The continuum in both archival and model spectra were normalized to unity using the Astropy \citep{Astropy} package specutils \citep{specutils} excluding the region 4075~\AA\ to  4125~\AA\ (near the strong H$\delta$ line at 4101~\AA).
Also shown in Figure~\ref{fig:spectrum_bandpass} is a model spectrum with $T_{\rm eff}$ = 9190 K, the effective temperature value from \cite{Jones_2015} (see Section \Ref{sec:Age}), which displays stronger lines, particularly in the wings on the H$\delta$ line, that are not present in the archival spectrum.

\begin{figure*}[t!]  
  \includegraphics[width=\linewidth]{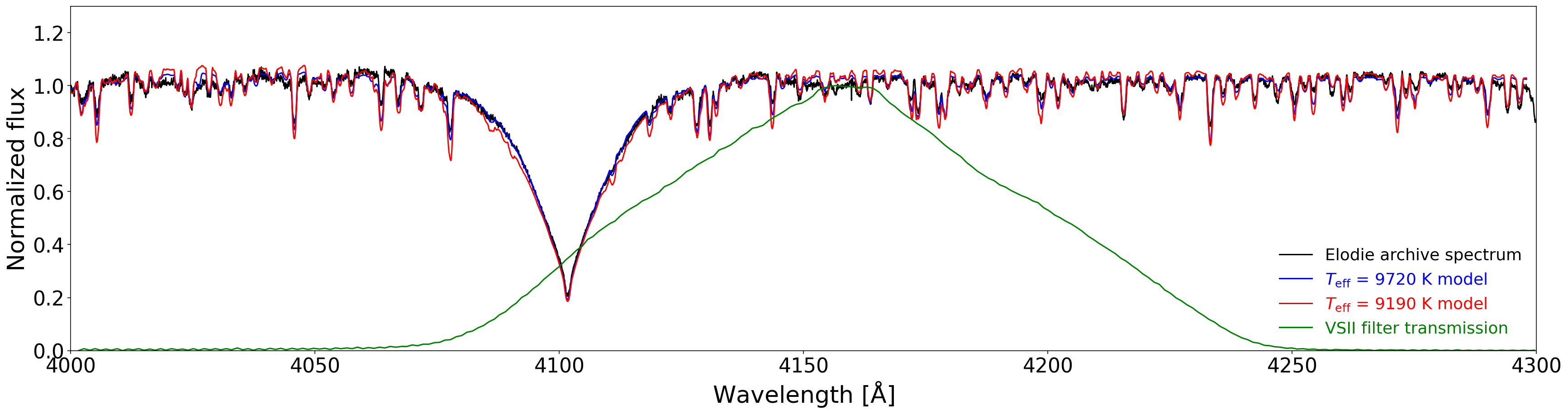}
  \caption{The measured VSII bandpass (in green) is shown in comparison to a normalized ELODIE archive 
  \citep{elodie}  spectrum of $\beta$ UMa (observation 20040309/0031, in black)
  and a normalized model spectrum (in blue) from the same model (mean effective temperature, $T_{\rm eff}$ = 9720 K,  mean surface gravity (in cgs units), $\log_{10}(g/{\rm cm\, s^{-2}}) = 3.93$, mass = 2.56 M$_\odot$, $v\sin i = 46$ $\rm km\, s^{-1}$)
  used to compute the synthetic visibilities.
A cooler model spectrum (in red) with $T_{\rm eff}$ = 9190 K is also shown.}
    \label{fig:spectrum_bandpass}
\end{figure*}
  
Setting $M_i$ in Equation~\ref{eq:chi2} to these synthetic visibilities, a best-fit diameter of $\theta_{\rm LD}=1.07\pm0.04$\,mas is found. 
This value along with previous angular diameter measurements and selected estimates for $\beta$ UMa are compiled in Table \ref{tab:Literature_Angular_Diameters}.

As a cross-check and also to check the effect of baseline variations during a run, we found that these numerical calculations are in excellent agreement with an analytic functional form introduced by~\cite{HB1974_LD} 
\begin{equation}
\label{eq:LimbDarkenedModel}
    V_{\rm LD}^2(\theta_{LD},b) = C_{\rm LD}\times\left(\frac{\alpha \frac{J_1(x)}{x} + \beta\sqrt{\frac{\pi}{2}}\frac{J_{3/2}(x)}{x^{3/2}}}{\frac{\alpha}{2}+\frac{\beta}{3}}\right)^2 ,
\end{equation}
where $\alpha=1-u_\lambda$, $\beta=u_\lambda$, $x=\pi b\, \theta_{\rm LD}/\lambda$.
The parameter $u_\lambda$ is the linear limb-darkening coefficient, which is wavelength dependent. 
Setting {$u_\lambda=0.45$} is consistent with our model atmosphere center-to-limb intensity profile in the VSII bandpass. 
This $u_\lambda$ value falls between values computed for the Johnson U and B bands for $T_{\rm eff}$ = 10000 K and $\log_{10}(g/{\rm cm\, s^{-2}}) = 4.00$ by \cite{LD_coefficient}.

\begin{figure}[t!]
    \centering{\includegraphics[width=0.8\linewidth]{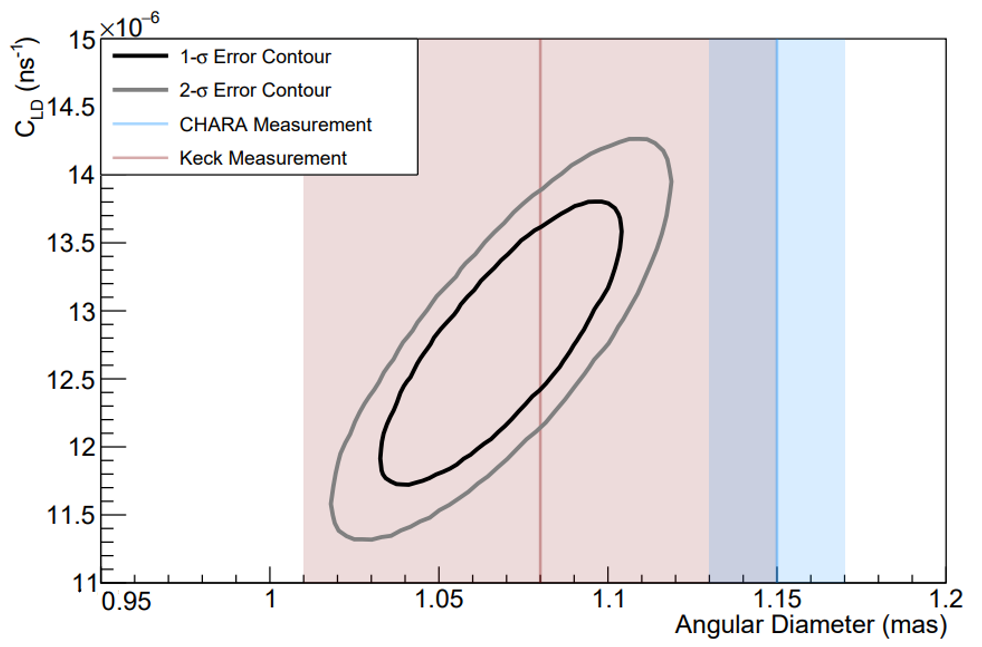}}
    \caption{Error contours for Chi-square minimization (equation~\ref{eq:chi2}) between the synthetic visibilities and the visibilities in Figure \ref{fig:UniformDiskVisibilites}, with limb-darkened measurements from the CHARA~\citep{Boyajian_2012} and Keck~\citep{Mennesson_2014} arrays shown as shaded regions.}
    \label{fig:ErrorContour}
\end{figure}

Replacing $M(\theta_D,b)$ with $V_{\rm LD}(\theta_{\rm LD},b)$ in Equations~\ref{eq:AveragingModels} and~\ref{eq:chi2}, we again find $\theta_{\rm LD}=1.07\pm0.04$\,mas.
Analysis-cut variations discussed in Section~\ref{sec:UniformDiskDiameterPrimary} result in an estimate for systematic uncertainty of 5\%.
Chi-square confidence levels of the fit are shown in Figure~\ref{fig:ErrorContour}, together with limb-darkened diameters measured with the CHARA and Keck arrays.

\section{The Age and Effective Temperature of $\beta$ UMa} \label{sec:Age}

We have derived fundamental stellar parameters for $\beta\,\rm UMa$ using the VSII limb-darkened angular diameter along with literature values for the bolometric flux and parallax, together with model stellar evolution tracks, see Table \ref{tab:stellar_parameters}.
Our stellar atmosphere models include a small degree of rotational distortion (0.5\%) consistent with the rotational broadening of the spectral lines shown in Figure \ref{fig:spectrum_bandpass}.
The corresponding small degree of pole-to-equator gravity darkening corresponds to a temperature difference of 30 K assuming the equatorial velocity is equal to the observed $v\sin i = 45\, {\rm km\, s^{-1}}$ \citep{rotation}, where $i = 90^\circ$.

\begin{deluxetable*}{lCl}[b]
\label{tab:stellar_parameters}
\tablecaption{Fundamental Stellar Parameters for $\beta$ UMa}
\tablecolumns{3}
\tablewidth{0pt}
\tablehead{
\colhead{Parameter} &
\colhead{Value} & 
\colhead{Reference}}
\startdata
Limb-darkened angular diameter, $\theta_{\rm LD}$ (mas)    &1.07\pm0.04$\pm$0.05   &This paper\\
Bolometric flux at Earth, $F_{\rm bol}$ (${\rm erg\,s^{-1}\, cm^{-2}}$)    &(340 \pm 7)\times10^{-8} &\cite{Boyajian_2012}\\
Effective temperature, $T_{\rm eff}$ (K)                 &9700\pm200\pm200 &derived, $\bigl[{4 F_{\rm bol}}/{\sigma \theta_{\rm LD}^2}\bigr]^{1/4}$ \\
Parallax, $\varpi$ (mas)                          &40.90\pm 0.16 &\citet{V07}\\
Radius, $R$ ($R_\sun$)                           & 2.81\pm0.11\pm0.13  &derived, $\theta_{LD}/2\varpi$ \\
Luminosity, $L$ ($L_\sun$)                       &63.5\pm 1.4  &derived, $4\pi F_{\rm bol}/\varpi^2$\\
Mass, $M$ ($M_\sun$)                            &2.56\pm0.03\pm0.02 & MIST tracks \citep{MIST0, MIST1}\\
$\log_{10}$ surface gravity, $\log g$ (${\rm cm\,s^{-2}}$)             &3.93\pm0.03 \pm 0.05 &derived, $g = GM/R^2$\\
Age (Myr)                                       &390\pm 29\pm 32  & MIST tracks \citep{MIST0, MIST1} \\
Projected rotational velocity, $v\sin\,i$ (${\rm km\,s^{-1}}$) &47\pm 3 &\citet{rotation}\\
\enddata
\tablecomments{Uncertainties propagated into the derived parameters include both statistical and systematic uncertainties in the limb-darkened angular diameter, except for the luminosity which is independent of the measured angular size of the star.}
\end{deluxetable*}

For slow rotators, long-baseline interferometry helps constrain stellar ages via the effective temperature, since the inferred luminosity depends only on the measured bolometric flux and the parallax, not the angular diameter.
Figure \ref{fig:MIST_tracks} shows mass and age constraints, using uncertainties on effective temperature and luminosity (see Table \ref{tab:stellar_parameters}), based on stellar evolutionary tracks from the \dataset[MESA Isochrones and Stellar Tracks (MIST) web interpolator]{http://waps.cfa.harvard.edu/MIST/interp_tracks.html}
\citep{MIST0,MIST1}. The model tracks have solar metallicity, with an
initial rotational velocity set to 40\% of the critical velocity.
The blue and orange tracks in Figure~\ref{fig:MIST_tracks} mark an inner 
stellar mass range, 2.535 to 2.585 $M_\sun$ 
(based on $T_{\rm eff} = 9700\pm 200$ K, statistical uncertainty only, see Table
\ref{tab:stellar_parameters}), and purple and red tracks mark
the outer stellar mass range, 2.520 to 2.605 $M_\sun$ 
(based on $T_{\rm eff} = 9700\pm 400$ K, adding statistical 
and systematic uncertainties, not in quadrature). The inner age range spans 357 to 414  Myr and the outer age range spans 322 to 443 Myr.

Our age estimate for $\beta$ UMa, $390 \pm 29 \pm 32$ Myr, is consistent
with  \cite{Jones_2015},  $408 \pm 6 $ Myr. Our younger
mean age comes from our higher mean effective temperature, 
$9700\pm 200\pm 200$ K, relative to $9190 \pm 56$ K. This cooler value for $T_{\rm eff}$ in \cite{Jones_2015}
is surprising given the value of 9377 $\pm$ 75 K from \cite{Boyajian_2012} which is consistent with the CHARA limb-darkened angular diameter (see Table \ref{tab:Literature_Angular_Diameters}) and the bolometric flux, even though both papers cite the same physical radius, 3.021$\pm$ 0.038  $R_\odot$. 
A model spectrum with $T_{\rm eff}$ = 9190 K, shown together with an observed spectrum in Figure \ref{fig:spectrum_bandpass},
appears to be too cool: the metal lines are all stronger, particularly in the wings of the H$\delta$ line. 
Also, a $T_{\rm eff}$ = 9190 K model has insufficient flux to match the archival absolute ultraviolet spectrophotometry shown in Figure \ref{fig:SED}. Likewise, a $T_{\rm eff}$ = 9377 K  model also has 
insufficient flux in the shorter-wavelength ultraviolet, while a $T_{\rm eff}$ = 9720 K is a 
better match to the full spectral energy distribution (SED), consistent with \cite{Swihart2017} who used spectrophotometry to fit an angular diameter, see Table \ref{tab:Literature_Angular_Diameters}.
A better match to the observed SED could likely be achieved from a full non-local 
thermodynamic equilibrium (non-LTE) model atmosphere, with metal abundances adjusted to more carefully match the high-resolution spectrum, beyond the scope of the present analysis.

The lower effective temperature values could possibly be due to a systematically high CHARA
angular diameter. As noted in \cite{Boyajian_2012}, CHARA Classic angular diameter measurements are systematically
higher by a factor of $1.06\pm0.06$\,mas compared to identical measurements by the Palomar Testbed Interferometer (PTI) (also see \cite{CHARA_vs_PTI}). We note, however, that $\beta$ UMa was never observed by PTI due to its small angular size, and so a direct comparison between CHARA and PTI measured diameter measurements is not available for this source. 

\begin{figure*}[h!]  
  \includegraphics[width=\linewidth]{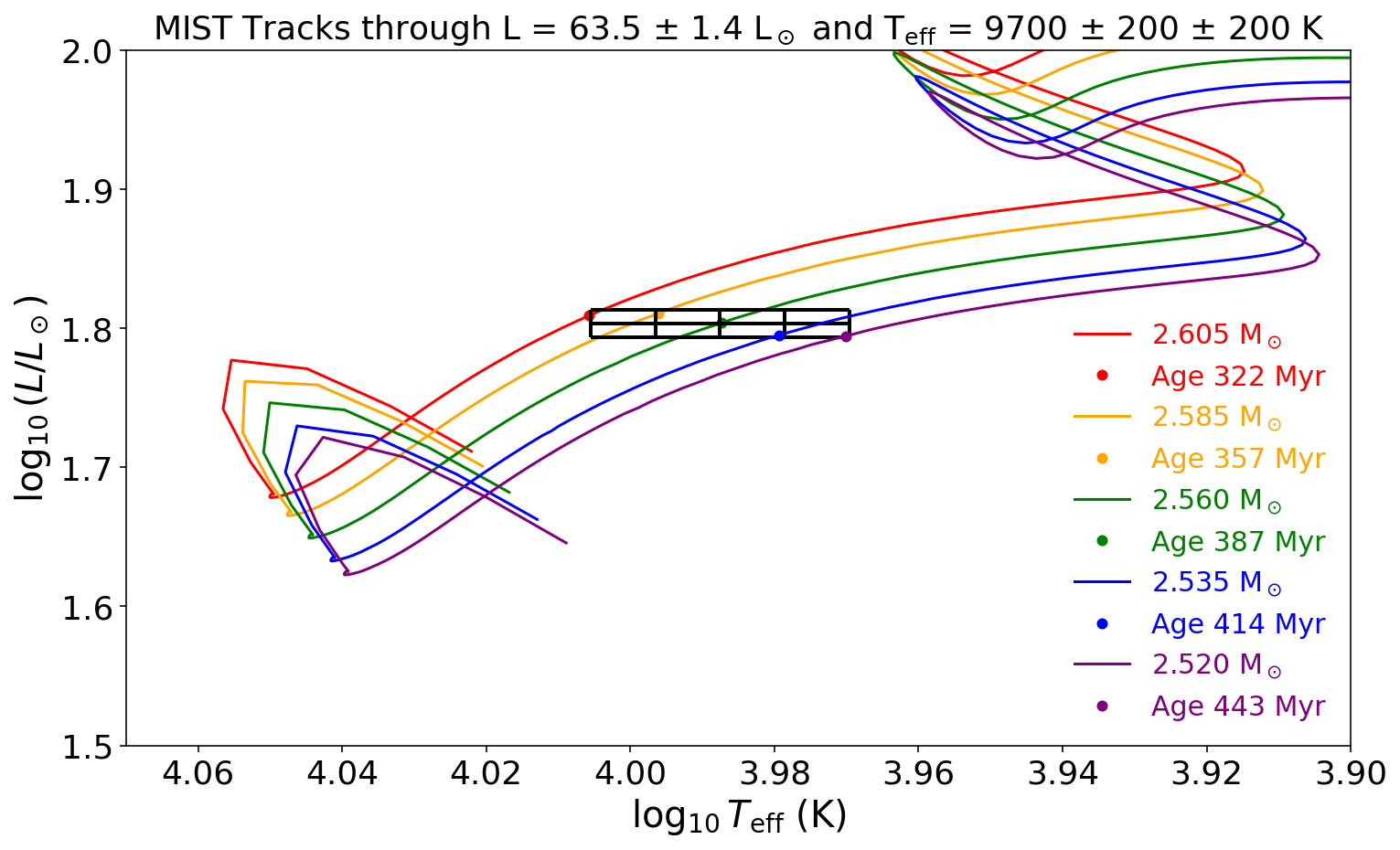}
  \caption{Model evolutionary tracks, from the 
  \dataset[MESA Isochrones and Stellar Tracks (MIST) web interpolator]{http://waps.cfa.harvard.edu/MIST/interp_tracks.html} \citep{MIST0,MIST1}, through constraints on the effective temperature and luminosity. These constraints are provided by  the VSII limb-darkened angular diameter, the bolometric flux,  and the parallax, see Table \ref{tab:stellar_parameters}.
  Model tracks have solar metallicities, with an initial rotational velocity set to 40\% of the critical velocity.}  
  \label{fig:MIST_tracks}
\end{figure*}

\begin{figure*}[h!]  
  \includegraphics[width=\linewidth]{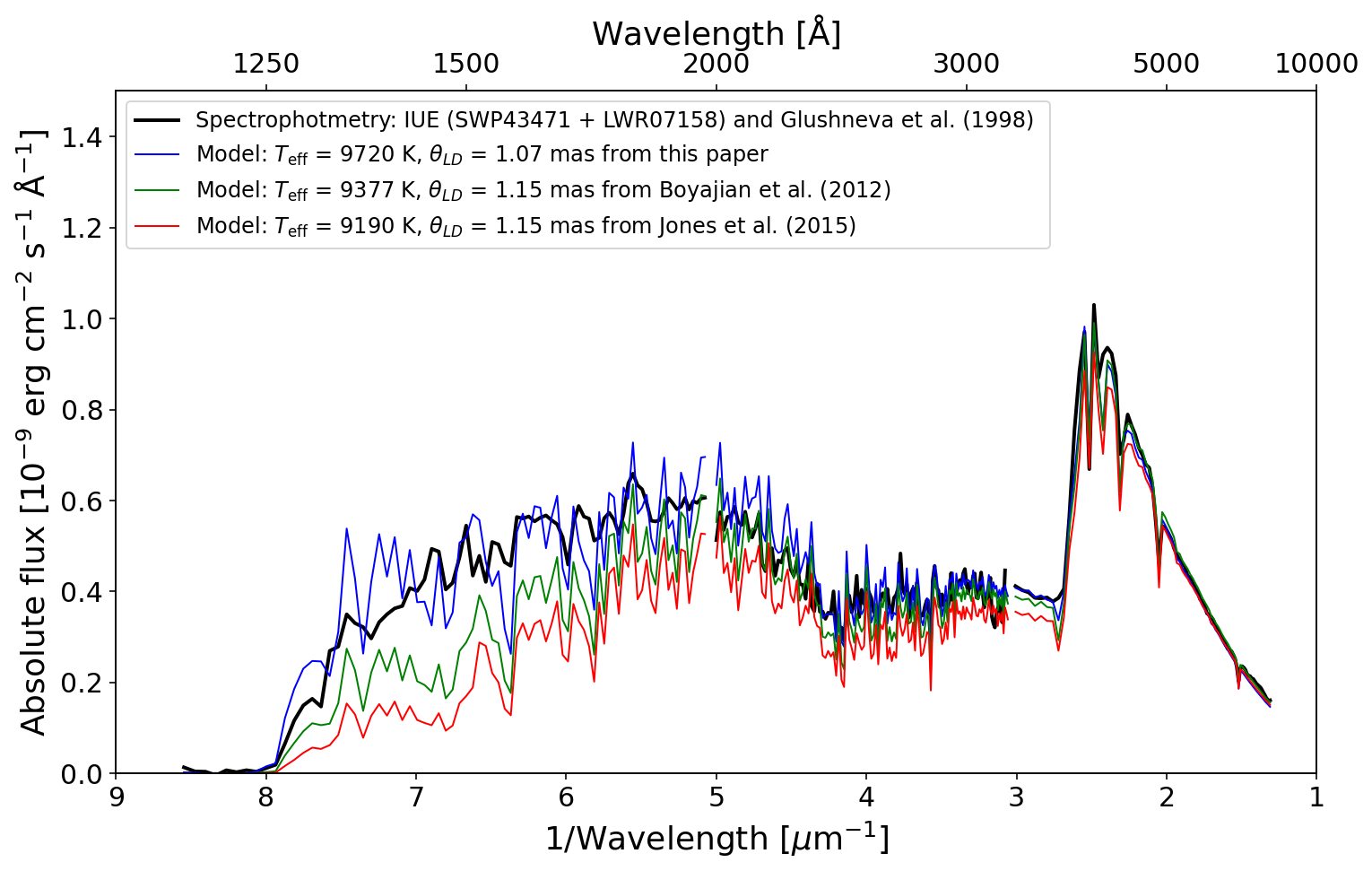}
  \caption{Ultraviolet (UV) spectrophotometry from the International Ultraviolet Explorer (IUE), specifically from the Short
  Wavelength Prime (SWP) and Long Wavelength Redundant (LWR) cameras, both with the large aperture, are shown together optical spectrophotometry
  from \cite{glushneva98}. For clarity, in the UV, both the IUE and model fluxes are binned to 1 nm.
  In the optical, the data and model fluxes are binned to 5 nm. Three model spectral energy distributions using
  the effective temperature and limb-darkened angular diameters from this paper (blue), 
  \cite{Boyajian_2012} (green) and \cite{Jones_2015} (red). These
  IUE data (never shown before in the literature) were retrieved from the MAST archive at \dataset[10.17909/jgah-0b96]{https://dx.doi.org/10.17909/jgah-0b96}.}
  \label{fig:SED}
\end{figure*}

\section{Summary and Future Work}\label{sec:discussion}
We have utilized an updated version of VSII data reduction algorithms to obtain a limb-darkened angular diameter of 1.07 $\pm$ 0.04 $\pm$ 0.05 for $\beta$ UMa, the first angular diameter measurement of this star at visual wavelengths (416 nm).  
The mean value of the uniform-disk angular diameter at 416 nm is smaller than diameters observed at longer wavelengths, consistent with the expectation of stronger limb darkening at shorter wavelengths (see Table \ref{tab:Literature_Angular_Diameters}).  
This measurement allowed us to derive fundamental parameters for $\beta$ UMa, listed in Table \ref{tab:stellar_parameters}.
A model atmosphere with the effective temperature based on our measured limb-darkened angular diameter and the measured bolometric flux provides a very good match to the observed high-resolution spectrum in the 416 nm band (Figure \ref{fig:spectrum_bandpass}) and is consistent with absolute spectrophotometry from 120 nm to 830 nm.
Our age estimate for $\beta$ UMa, $390 \pm 29 \pm 32$ Myr (see Figure \ref{fig:MIST_tracks}), agrees with that from \cite{Jones_2015},  408 $\pm$ 6 Myr. 

In the measurement presented here, the statistical uncertainty is similar to the systematic uncertainty. Accumulating more observation statistics with additional exposure would do little to provide more precise input to stellar models.
The VERITAS collaboration is in the process of implementing several upgrades to the SII system, to reduce the systematic uncertainties in future observations.

As discussed extensively above, the uncertainty in the relative time delay between the recorded data at each telescope acquisition system severely complicates the measurement of $g^{2}$ under low signal-to-noise conditions. 
This limits the ability to accurately measure the angular diameters of dim targets and cooler stars, and also limits the information that can be extracted at longer baselines that present low signal-to-noise measurements of the squared visibility. 
In the future, the timing uncertainty will be eliminated by injecting sub-nanosecond synchronized square pulse signals of ~10 ns duration into each SII telescope data stream at a rate of 1 Hz. 
Comparing the arrival time of these calibration pulses in each telescope will constrain telescope relative time delays to substantially less than the 4 ns sampling time. 

Another fundamental limitation is the uncertainty in the amplitude (equivalently the integral) of the $g^{(2*)}$ peak at zero baseline.
The zero baseline correlation (ZBC) is currently left as a free parameter ($C_{\rm UD}$ and $C_{\rm LD}$ in equations~\ref{eq:UniformDiskModel} and~\ref{eq:LimbDarkenedModel}, respectively) in our angular diameter fits. 
A direct measurement of the ZBC would enable model-independent estimates of the squared visibility. 
An empirical measurement of the ZBC can be performed by carefully modifying the focal plane instrumentation with optics that enable the light to be divided into two detectors. 
We estimate that constraining the ZBC to better than 5\% would approximately halve the current uncertainty on the angular diameter.

Statistical uncertainties will be reduced by maximizing the collected starlight delivered to each SII photomultiplier tube. 
First, the recoating of the VERITAS telescope mirror facets has begun and will continue into 2024. 
However, this process had not yet started when the observations presented in this paper were taken. The recoating will increase overall mirror reflectivity at the SII bandpass (416 nm) by up to 50\%. 
Second, work is underway to install a tracking correction system that will keep the focal plane image of the observed star continually centered on the SII PMT. 
The correction system uses a matrix of plastic fiber optics surrounding the SII PMT to look for small misalignments between the star image on the focal plane and the SII PMT. 
A CCD camera readout of the fiber optic light intensities is used to calculate the position of the star image on the focal plane and send micro-adjustment instructions to the telescope tracking software to recenter the stellar image on the PMT. 
The improved alignment between the focal-plane image of the star and the SII PMT will increase the intensity and stability of the collected starlight, thereby improving the signal-to-noise ratio of the observation, and reducing potential systematics.  

Finally, an improved high-speed software correlator has been successfully demonstrated in the laboratory. 
Using a multi-threaded correlation code on 80 high-speed CPU cores, the new correlator will enable nearly real-time calculation of intensity correlations from all 6 telescope pairs simultaneously. 
The new correlator will be able to perform the correlation analysis over a wider range of time-lags, thereby improving the measurement of the background noise levels in the correlation functions. 
The new correlator system is also more easily configurable to test new algorithms by comparison to the currently used FPGA correlator. 
The new correlator was installed at VSII in Summer 2023. 
The new software-based correlator allows for real-time monitoring of SII data quality, thereby enabling real-time adjustments that reduce systematic errors in future measurements of stellar radii.

\newpage
\appendix
\section{Appendix: Correlation Peak Integrals and Model Dependent Squared Visibilities}
Table~\ref{tab:SquaredVisibilities} lists the measured areas of correlation peaks in $g^{2}_{AB}(\tau)$ shown in Figure~\ref{fig:UniformDiskVisibilites}.
The third column is simply the second column divided by $C_{\rm UD}$ to give a value of the squared visibilities given our uniform disk model.
\end{nolinenumbers}

\startlongtable
\begin{deluxetable*}{ccc}
\label{tab:SquaredVisibilities2}
\tablecaption{VSII squared visibilities for $\beta$ UMa}
\tablecolumns{3}
\tablewidth{0pt}
\tablehead{
\colhead{Average Baseline} \vspace{30cm} &
\colhead{$g^{2}_{AB}(\tau)-1$ Area} &
\colhead{Model Dependent}\\  
\colhead{(meters)} &
\colhead{($10^{-6}$\,ns)} &
\colhead{Squared Visibility}
}
\decimals
\startdata
 35.4 &10.0 $\pm$ 1.9 & 0.81 $\pm$ 0.15 \\
 36.3 & 7.6 $\pm$ 1.5 & 0.60 $\pm$ 0.12 \\
 38.3 & 7.3 $\pm$ 1.0 & 0.57 $\pm$ 0.08 \\
 39.3 & 6.2 $\pm$ 1.0 & 0.50 $\pm$ 0.08 \\
 39.4 & 5.9 $\pm$ 1.1 & 0.45 $\pm$ 0.09 \\
 43.1 & 8.5 $\pm$ 1.3 & 0.67 $\pm$ 0.10 \\
 44.4 & 8.6 $\pm$ 1.2 & 0.68 $\pm$ 0.10 \\
 47.6 & 4.7 $\pm$ 1.0 & 0.37 $\pm$ 0.08 \\
 51.1 & 5.7 $\pm$ 1.1 & 0.45 $\pm$ 0.08 \\
 53.2 & 4.1 $\pm$ 0.9 & 0.32 $\pm$ 0.07 \\
 56.1 & 7.5 $\pm$ 1.6 & 0.59 $\pm$ 0.12 \\
 58.7 & 1.9 $\pm$ 0.6 & 0.15 $\pm$ 0.05 \\
 59.5 & 3.6 $\pm$ 1.1 & 0.28 $\pm$ 0.08 \\
 60.3 & 4.4 $\pm$ 2.3 & 0.35 $\pm$ 0.18 \\
 67.6 & 3.1 $\pm$ 0.8 & 0.24 $\pm$ 0.06 \\
 68.2 & 2.5 $\pm$ 0.6 & 0.20 $\pm$ 0.05 \\
 68.9 & 2.1 $\pm$ 1.5 & 0.16 $\pm$ 0.12 \\
 71.6 & 2.4 $\pm$ 0.7 & 0.19 $\pm$ 0.06 \\
 73.6 & 3.4 $\pm$ 0.7 & 0.27 $\pm$ 0.05 \\
 74.2 & 0.3 $\pm$ 0.6 & 0.03 $\pm$ 0.04 \\
 75.6 & 2.0 $\pm$ 0.6 & 0.15 $\pm$ 0.05 \\
 78.6 &-0.7 $\pm$ 0.6 &-0.05 $\pm$ 0.04 \\
 79.0 &-0.7 $\pm$ 0.5 &-0.06 $\pm$ 0.04 \\
 81.2 & 2.1 $\pm$ 0.9 & 0.17 $\pm$ 0.07 \\
 83.6 & 1.4 $\pm$ 0.9 & 0.11 $\pm$ 0.07 \\
 84.4 & 2.0 $\pm$ 1.1 & 0.16 $\pm$ 0.09 \\
 89.5 &-0.1 $\pm$ 0.4 &-0.01 $\pm$ 0.03 \\
 97.6 & 0.0 $\pm$ 0.6 & 0.00 $\pm$ 0.05 \\
\enddata
\tablecomments{The $g^{2}_{AB}(\tau)-1$ and model dependent squared visibilities for the baselines recorded at VERITAS. These values are shown in Figure \ref{fig:UniformDiskVisibilites}.} 
\label{tab:SquaredVisibilities}
\end{deluxetable*}

\newpage
\nolinenumbers
\section{Appendix: Secondary analysis} \label{sec:secondary}

The primary analysis results were verified through an independent secondary analysis employing a Monte Carlo Markov Chain (MCMC) Bayesian analysis implemented in the Cobaya software package (\cite{cobaya2021}). Briefly, the secondary analysis models the SII visibility signal using the analysis methodology outlined in \citep{Davis2022MasterThesis}, augmented by the inclusion of the observed statistical noise associated with non-correlated stellar light, night sky background fluctuations and broadband electronic noise (not including narrowband radio-frequency pickup).  The statistical noise model was verified through Monte Carlo simulation (described in Appendix \ref{appen:secondaryModel}). 

The secondary analysis employs a simple low bandpass digital filter to eliminate observed narrowband radio frequency contamination from individual raw correlation functions. The analysis then corrects the 
value of $g^{(2*)}_{ab}$ for night sky background effects to obtain $g^{(2*}_{ab}$,
similar to the primary analysis equation \ref{eq:DarkRunFactor}. The secondary analysis then fits the subsequent visibility curve  (including both visibility signal and statistical noise) to the analytical limb-darkened angular diameter model given in equation \ref{eq:LimbDarkenedModel}. Importantly, the secondary analysis uses the statistical noise model to correct for noise-induced uncertainty in the location of the correlation coherence peaks in the two-telescope correlation functions, as discussed in Section \ref{sec:QuantifyingVisibility}.  The statistical noise model was used to estimate the magnitude of systematic biases of the fitted parameters for both the primary and secondary analyses.

\begin{figure}
  \includegraphics[width=0.9\linewidth]{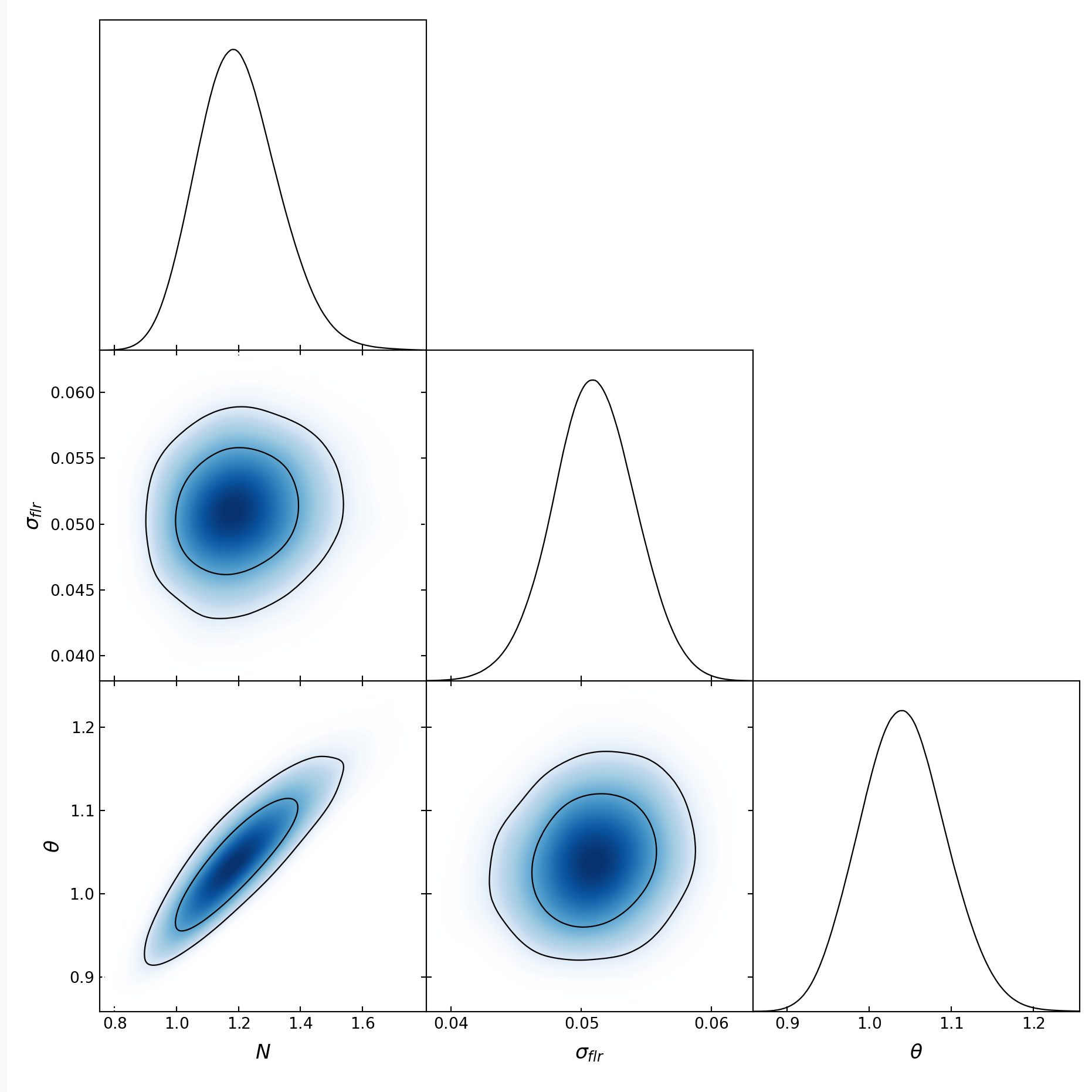}
  \caption{The 95\% and 68\% parameter constraint contours for a limb darkened model, $\theta$ the angular diameter in units of mas, $N$ is a constant multiplier (normalization parameter in units of $10^{-6}$) of the limb darkened model given in equation \ref{eq:LimbDarkenedModel}, with a noise parameter included in the modeling $\sigma_{flr}$ defined in appendix \ref{appen:secondaryModel}. The mean of the limb darkened angular diameter parameter chain is 1.043 $\pm$ 0.049 mas. The top plot of each column shows a visual of the 1D posterior probability of the parameter at the bottom of the same column. 
  }
  \label{fig:MCMCResults}
\end{figure}

The results of the secondary analysis  are presented in Figure \ref{fig:MCMCResults}. 
The secondary analysis calculates a limb-darkened stellar diameter of 1.04 $\pm$ 0.05 mas (stat), consistent with the primary analysis. The secondary analysis statistics are presented in Table \ref{tab:mcmcRes}. As the secondary analysis fits both the visibility curve model and statistical noise model simultaneously, biases due to uncertainties in the timing location of the $g^2$ peak are implicitly included. The secondary analysis therefore eliminates the potential systematic biases in the primary analysis associated with the the handling of visibility points at low statistical significance, as described in Section \ref{sec:QuantifyingVisibility}. The estimated uncertainty on the measured angular diameter in the secondary analysis is increased by an excess factor of 1.5 to account for non-linearities in the statistical model and correlated parameters. This multiplicative factor was determined using Monte Carlo simulations.  

There are several important caveats to consider regarding the secondary analysis. First, we have assumed that statistical noise within the correlation function is normally distributed (broadband) and the narrow-band digital filter has eliminated individual narrow-band radio frequency lines. The narrow-band digital filter may introduce non-Gaussian components to the broadband statistical noise. For the observed statistical uncertainties in $\theta_{\rm LD}$ ($\sim$4\%), we have determined that these systematic effects are insignificant. However, these effects may begin to add a significant systematic contribution when the statistical uncertainty in $\theta_{\rm LD}$ approaches 1\%.  The average uncertainty ($\sigma_{\mathrm{flr}}$) has been observed to be constant. A potential systematic bias induced by the variability of $\sigma_{\mathrm{flr}}$ is only quantifiable through extensive simulation. Finally, our analysis assumes that a uniform or limb-darkened circular disk well approximates the source image of the  $\beta$ UMa, which is known to be a slow rotator. 

We have performed multiple simulations to explore how a combination of these effects might bias the results of the secondary analysis. For reasonable parameter ranges, the simulations indicate that the reconstructed angular diameter is biased by less than $\sim$2\% of the input angular diameter. Consequently, the results of the primary and secondary analyses are found to be in excellent agreement. Even for more extreme parameter ranges, the secondary analysis results remain in agreement with the primary analysis to within $\pm 1\sigma$.


\label{sec:mesRes}

\subsection{The statistical noise model used in the secondary analysis}
\label{appen:secondaryModel}
Due to uncertainties in the expected time location of  $g^{(2)}_{ab}$  peak in the two-telescope correlation functions, its estimated location is determined by analysis of the correlation function. At low signal-to-noise levels, the location determination can misidentify the location of the $g^{(2)}_{ab}$ peak in the correlation function, and potentially bias the measurement of its amplitude.


We use a two-component statistical noise model to calculate the statistical fluctuations in each bin of a measured correlation function. This model will calculate the fluctuations in each correlation function bin, and will therefore characterize the fluctuations that potentially change the location and amplitude of the  $g^{(2)}_{ab}$ peak.  One component of the model contains the photostatistical fluctuations in the bins containing the $g^{(2)}_{ab}$ peak, whereas the other component includes broadband statistical noise fluctuations in every bin drawn from a distribution of normal Gaussian fluctuations around a defined average noise level $\sigma_{flr}$. This second component represents the photostatistical fluctuations in non-correlated stellar light, background sky noise, and additional fluctuations due to broadband electronic noise. The statistical model does not include any narrowband radio frequency line interference noise.

We model the statistical fluctuations in the correlation function  by determining the joint expectation value of these two random variables in each correlation function bin.  The statistical noise model calculates what is expected to be measured, on average, if one were to search for the maximum correlation peak within a given correlation function with $n$ time bins, assuming an average noise in each bin of $\sigma_{flr}$, alongside a given expectation value of $g^{(2)}_{ab}$ in the correlation function from a squared visibility model $\operatorname{E}_{\mathrm{V^2}}$.

We start with the random variable for the correlation function bins lacking any $g^{(2)}_{ab}$ signal. Borrowing from order statistics theory, the maximum of the cumulative distribution function $CDF_{\mathrm{MAX}} (X,n)$ for a  random variable $X$ and $n$ bins is related to the cumulative distribution function  $CDF(X,n)$  for a set of $n$ random variables  $X_1, X_2 ... X_n$:

\begin{equation}
    CDF_{\mathrm{MAX}} (X,n) = [CDF(X)]^n.
\end{equation}

The probability density function ($PDF$) is the derivative of the $CDF$. Taking the derivative of the $CDF$ using the chain rule,  we obtain
\begin{equation}
    PDF_{\mathrm{MAX}} (X,n) = n [CDF(X)]^{n-1} PDF(X).
\end{equation}

The standard normal distribution is defined as
\begin{equation}
    \phi(X) = e^{-X^2 /2}.
\end{equation}
The standard cumulative distribution is defined as
\begin{equation}
    \Phi(X) = 0.5(1+\operatorname{erf}(X/\sqrt{2})),
\end{equation}
where $\operatorname{erf}$ is the error function
\begin{equation}
    \operatorname{erf}(X) = \frac{2}{\sqrt{\pi}} \int_{0}^{X} e^{-t^2} dt.
\end{equation}

Combining the equation for the standard normal distribution and standard cumulative distribution, we derive
\begin{equation}
    PDF_{MAX \phi} (X,\sigma_{\mathrm{flr}},n) = n \left[\Phi(\frac{X}{\sigma_{\mathrm{flr}}})\right]^{n-1} \phi(\frac{X}{\sigma_{\mathrm{flr}}}),
\end{equation}

where $\sigma_{\mathrm{flr}}$ is the standard deviation of the random set $X_1, X_2 ... X_n$. 



The standard deviation on the expectation value of the maximum is calculated as 
\begin{equation}
    \sigma_{\mathrm{max}}(\sigma_{\mathrm{flr}}) = 
    \sqrt{\operatorname{E}_{\mathrm{max}}^2(\sigma_{\mathrm{flr}}) - 
    \left(\operatorname{E}_{\mathrm{max}}(\sigma_{\mathrm{flr}})\right)^2 },
\end{equation}


where the expectation value of the maximum $\operatorname{E}_{\mathrm{max}}(\sigma_{\mathrm{flr}})$ is defined by 
\begin{equation}
    \operatorname{E}_{\mathrm{max}}(\sigma_{\mathrm{flr}}) = 
    \int_{-\infty}^{\infty} \frac{xn}{\sqrt{2\pi}\sigma_{\mathrm{flr}}} 
    \phi\left(\frac{x}{\sigma_{\mathrm{flr}}}\right)
    \left[\Phi\left(\frac{x}{\sigma_{\mathrm{flr}}}\right)\right]^{n-1} dx,
\end{equation}


and $\operatorname{E}_{\mathrm{max}}^2(\sigma_{\mathrm{flr}})$ is given by
\begin{equation}
    \operatorname{E}_{\mathrm{max}}^2(\sigma_{\mathrm{flr}}) = 
    \int_{-\infty}^{\infty} \frac{x^2n}{\sqrt{2\pi}\sigma_{\mathrm{flr}}} 
    \phi\left(\frac{x}{\sigma_{\mathrm{flr}}}\right) 
    \left[\Phi\left(\frac{x}{\sigma_{\mathrm{flr}}}\right)\right]^{n-1} dx.
\end{equation}

We can now approximate the off-peak correlation function bins (i.e. with no $g^{(2)}_{ab}$ signal)  as a single random variable with the expectation value $\operatorname{E}_{\mathrm{max}}$ and standard deviation $\sigma_{\mathrm{max}}$. 

Next, to approximate the correlation function bins within the $g^{(2)}_{ab}$ peak, we take the expectation value of the squared visibility as $\operatorname{E}_\mathrm{V^2}$ with a variance of 0. Taking from \cite{maxOfTwoGaus}, we then plug in all the necessary variables ($\operatorname{E}_{\mathrm{max}}$, $\sigma_{\mathrm{max}}$, and $\operatorname{E}_\mathrm{V^2}$) to model the expectation value of two independent, normally distributed random variables as
\begin{equation}
\begin{split}
    \label{eq:expectationV2}
    \operatorname{E}(\operatorname{E}_\mathrm{V^2}, \sigma_{\mathrm{flr}}, \operatorname{E}_{\mathrm{max}})_{g^2} \approx  
    \operatorname{E}_{\mathrm{V^2}} \Phi\left(\frac{\operatorname{E}_{\mathrm{V^2}} - \operatorname{E}_{\mathrm{max}}}{\sigma_{\mathrm{max}}(\sigma_{\mathrm{flr}})}\right) + 
    \operatorname{E}_{\mathrm{max}} \Phi\left(\frac{\operatorname{E}_{\mathrm{max}} - \operatorname{E}_\mathrm{V^2}}{\sigma_{\mathrm{max}}(\sigma_{\mathrm{flr}})}\right) + 
    \sigma_{\mathrm{max}}(\sigma_{\mathrm{flr}}) \phi\left(\frac{\operatorname{E}_\mathrm{V^2} - \operatorname{E}_{\mathrm{max}}}{1.5\sigma_{\mathrm{max}}(\sigma_{\mathrm{flr}})}\right),
\end{split}
\end{equation}

where the $1.5$ factor in the third term is a numerical correction derived through simulations with the uniform disk model. Monte Carlo simulations have shown that including this statistical correction to the visibility model results in residual bias on of order $<0.1$\% for the reconstructed angular diameter of a uniform disk model. 

%

\begin{deluxetable*}{||c|c|c|c|c||}[h!]
\label{tab:mcmcRes}
\tablecaption{Parameter Statistics Secondary Analysis}
\tablecolumns{5}
\tablewidth{0pt}
\tablehead{
\colhead{} &
\colhead{Mean} &
\colhead{STD} &
\colhead{Lower 68\%}&
\colhead{Upper 68\%}}
\decimals
\startdata
  $\theta$ & 1.04  & 0.05 & 0.99 & 1.09 \\
  \hline
  $N$ & 1.26 & 0.14 & 1.11 & 1.39 \\
  \hline
  $\sigma_{\mathrm{flr}}$ & 0.051 & 0.0031 & 0.048 & 0.054 \\
  \hline
  --- & Lower 95\% & Upper 95\% & Lower 99\% & Upper 99\% \\ 
  \hline
  $\theta$ & 0.95  & 1.15 & 0.92 & 1.18 \\
  \hline
  $N$ & 1.00  & 1.54 & 0.95 & 1.67 \\
  \hline
  $\sigma_{\mathrm{flr}}$ & 0.045 & 0.057 & 0.043 & 0.059 \\
  \hline
\enddata
\tablecomments{Parameter chain mean, parameter chain standard deviations (STD), and 68\%, 95\%, and 99\% intervals for a limb darkened visibility model. This includes the limb darkened angular diameter ($\theta$ mas), the normalization ($N \times 10^{-6}$), and the noise of the VSII instrument ($\sigma_{flr} \times 10^{-6}$).}
\end{deluxetable*}

\begin{figure*}[h!]
  \begin{center}
  \includegraphics[width=.95\linewidth]  {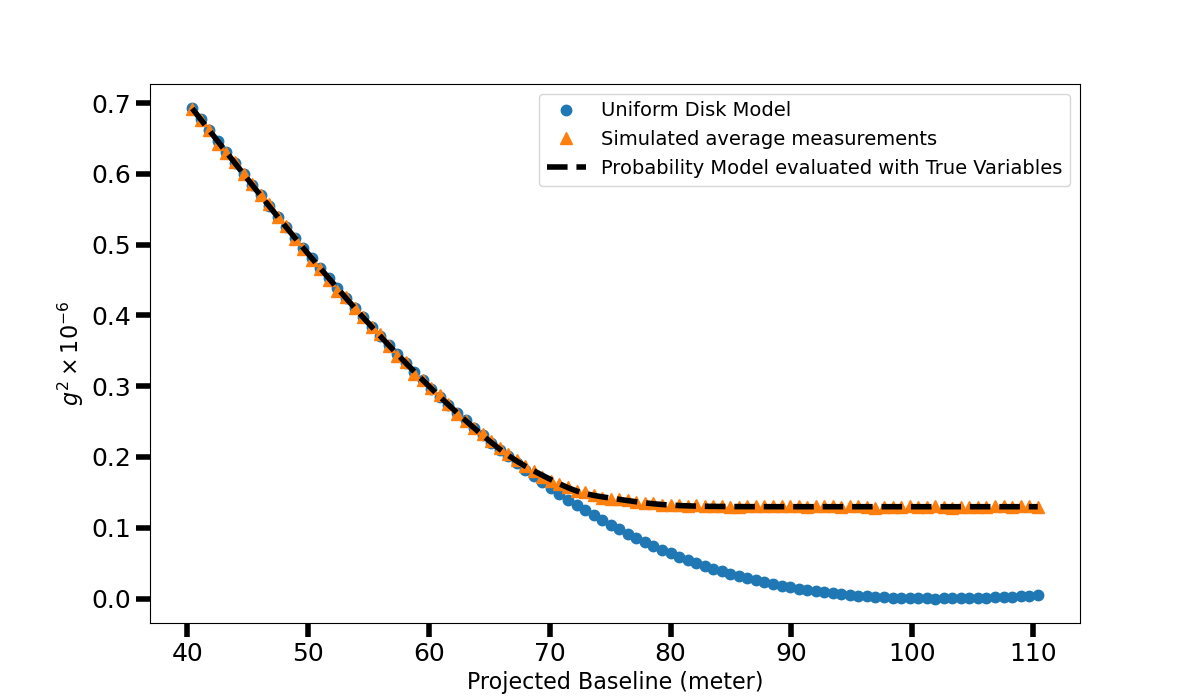}
  \end{center}
  \caption{An example simulation of the measurement of the maximum squared visibility as a function of projected baseline. The dashed line is the evaluation of the statistical model as given in equation \ref{eq:expectationV2} using $\sigma_{\mathrm{flr}}$ = 0.05, a normalization of 1.27 and an angular diameter of 1.2065 mas. Each triangle point is the average measured squared visibility of correlation function simulations (a simulated $g^2$ signal with normally distributed noise injected across 128 bins) as a function of the projected baseline. The blue circles are the underlying squared visibility model. All three input parameters are successfully measured to within 0.1\% using the scipy curve\_fit function \citep{2020SciPy-NMeth}.}
  \label{fig:statModelSim}
\end{figure*}

\newpage
\section*{Acknowledgements}
Both observations and analysis of data for the VERITAS Observatory are supported by grants from the US Department of Energy Office of Science, the US National Science Foundation, the Smithsonian Institution, and NSERC in Canada. 
The authors gratefully acknowledge support under NSF Grants \#AST 1806262 and \#PHY 2117641 for the fabrication and commissioning of the VSII instrumentation. 
We acknowledge the excellent work of the technical support staff at the Fred Lawrence Whipple Observatory and the collaborating institutions in the construction and operation of the VERITAS and VSII instruments. 
We acknowledge the support of the Ohio Supercomputer Center.  
This work made use of the Cray CS400 High Performance Cluster ``Vega'' at Embry-Riddle Aeronautical University (ERAU).  
J.~L. Sackrider was supported by ERAU Summer Undergraduate Research Fellowship. 
This research has made use of the VizieR catalogue access tool, CDS,  Strasbourg, France (DOI : 10.26093/cds/vizier). 
The original description of the VizieR service was published in 2000, A\&AS 143, 23. 
This research used observations made with the IUE mission, obtained from the MAST data archive at the Space Telescope Science Institute, which is operated by the Association of Universities for Research in Astronomy, Inc., under NASA contract NAS 5–26555.

\software{PHOENIX (version 18.07.03C)~\citep{H99},
Astropy~\citep{Astropy},
specutils~\citep{specutils},
Minuit~\citep{James:1975dr},
root~\citep{Brun:1997pa},
Cobaya ~\citep{cobaya2021}, 
SciPy ~\citep{2020SciPy-NMeth}.}

\bibliography{references}{}
\bibliographystyle{aasjournal}

\end{document}